\documentclass[fleqn,12pt,twoside]{article}
\usepackage{espcrc1}

\usepackage{epsfig}

\newcommand{\AmS}{{\protect\the\textfont2
  A\kern-.1667em\lower.5ex\hbox{M}\kern-.125emS}}

\title{J/$\psi$ production and suppression in nuclear collisions}

\author{Jianwei Qiu, James P. Vary, and Xiaofei Zhang \\
        Department of Physics and Astronomy, Iowa State University\\
        Ames, Iowa 50011, U. S. A.}
       
\begin{document}

\maketitle

\begin{abstract}
In terms of a new QCD factorization formula for J/$\psi$
production, we calculate the J/$\psi$ suppression in nuclear
collisions by including the multiple scattering between the
pre-J/$\psi$ partonic states and the nuclear medium.  We find
agreement with all data on J/$\psi$ suppression in hadron-nucleus and
nucleus-nucleus collisions, except a couple of points (the ``second
drop'') at the highest $E_T$ bins of the new NA50 data.
\end{abstract}

\section{J/$\psi$ Production}

J/$\psi$ suppression in relativistic heavy ion collisions was
suggested as a potential signal of the quark-gluon plasma
\cite{MS-jpsi}.  In order to study the J/$\psi$ suppression
quantitatively, we need to understand how the J/$\psi$ mesons are
produced, and the reliability of QCD calculations for J/$\psi$
production.

In order to produce a J/$\psi$ meson in hadronic collisions, 
the energy exchanges in the collisions have to be larger
than the J/$\psi$ mass ($M_{{\rm J/}\psi}$) or the invariant
mass of the produced quark pair 
($m_{c\bar{c}} \ge 2M_c \sim 3$~GeV).  
The $c\bar{c}$ pairs should be produced at a distance scale
$r_{\rm H} \le 1/2M_c \sim 1/15$~fm.  Since this is much
smaller than the size of a J/$\psi$ and 
the energy exchanges are much larger than the nonperturbative
momentum scales in the J/$\psi$ wave functions, the J/$\psi$ is
unlikely to be formed at the collision point. Instead, the J/$\psi$
meson should be formed after some resonance interactions (or coherent
soft gluon interactions between the charm and anticharm quarks).  
Thus, the transformation from a pre-J/$\psi$ $c\bar{c}$ pair to a
physical J/$\psi$ occurs over several Fermi \cite{BM-jpsi}.
During this transformation,
the produced $c\bar{c}$ pairs can also radiate gluons and
have interactions with spectators.  Due to
gluon radiation and the spectator interactions, it is possible for
the $c\bar{c}$ states, even with different quantum numbers,
to become a J/$\psi$ meson. 

However, because of the large energy exchange, the spectators'
roles in the hard collisions are suppressed by an extra
factor $\rho_N L/[(2M_c)^2+q_T^2]$ with the medium density $\rho_N$
and medium length $L$.  If the transverse momentum $q_T$ is large
enough, the spectator interactions are strongly suppressed, and the
cross sections for hadronic J/$\psi$ production can be factorized
\cite{QS-jpsi}, 
\begin{equation}
 \frac{d\sigma_{AB\rightarrow {\rm J/}\psi}}{dy d^2q_T} 
\approx \sum_{a,b}
 \int dx \phi_{a/A}(x)\int dx' \phi_{b/B}(x')
 \frac{d\hat{\sigma}_{ab\rightarrow{\rm J/}\psi}}{dy d^2q_T}(x,x')
\label{eq1}
\end{equation}
where $\sum_{ab}$ run over all parton flavors and the $\phi(x)$ are
the normal parton distributions.  The $\hat{\sigma}_{ab\rightarrow{\rm
J/}\psi}(x,x')$ represents the cross section for two partons
to produce a J/$\psi$.  Without the nuclear medium, $L$ is of the
order of the hadron radius ($R\sim 1$~fm).  

Because the production of the pre-J/$\psi$ $c\bar{c}$ states and the
transformation to the J/$\psi$ take place
at very different distance scales, quantum interference between
these two stages is suppressed by the ratio of these two distance
scales.  Therefore, the partonic cross sections for the J/$\psi$
production in Eq.~(\ref{eq1}) can be further factorized into 
two stages: (1) short-distance production of the pre-J/$\psi$ partonic
states, and (2) long-distance transition from the 
partonic states to the physical J/$\psi$ mesons \cite{QS-jpsi}, 
\begin{equation}
 \frac{d\hat{\sigma}_{ab\rightarrow{\rm J/}\psi}}{dy d^2q_T}
\approx
\sum_{[c\bar{c}]} \int \frac{dz}{z^2}\int dm_{c\bar{c}}^2 \left(
 \frac{d\hat{\sigma}_{ab\rightarrow [c\bar{c}]}}
      {dm_{c\bar{c}}^2 dy d^2p_{c\bar{c}_T}} \right)
 D_{[c\bar{c}]\rightarrow{\rm J/}\psi}(z;m_{c\bar{c}}^2)
\label{eq2}
\end{equation}
where $\sum_{[c\bar{c}]}$ sums over all $c\bar{c}$ states, 
$\vec{p}_{c\bar{c}}=\vec{P}_{{\rm J/}\psi}/z$ and $m_{c\bar{c}}^2 \ll
|\vec{p}_{c\bar{c}}|^2$.  The $D_{[c\bar{c}]}$ represent transition
probabilities from pre-J/$\psi$ $c\bar{c}$ states to J/$\psi$ 
mesons, and are defined by matrix elements of non-local
operators \cite{QS-jpsi}.  The differences between our approach and
other models of J/$\psi$ production can be cast as differences in 
$D_{[c\bar{c}]}$.

In QCD perturbation theory, the $D_{[c\bar{c}]}$ include all soft
resonance interactions and gluon radiations.  Similar to the
parton-to-hadron fragmentation functions, the $D_{[c\bar{c}]}$ depend
on nonperturbative physics.  Thus, the predictive power of the
factorized formula relies on the 
universality of these transition distributions \cite{QS-jpsi}.  

The non-relativistic QCD (NRQCD) model \cite{NRQCD-jpsi} and the color
evaporation (CE) model \cite{CE-jpsi} correspond to two diverse 
assumptions for the $D_{[c\bar{c}]}$.
The NRQCD model assumes that the $D_{[c\bar{c}]}$ are   
steeply falling functions of the invariant masses of the 
states, and sensitive to the pairs' quantum numbers (such as color and 
spin).  We can represent the leading contributions of this model by
expanding the partonic cross section $\hat{\sigma}_{ab\rightarrow
[c\bar{c}]}$ in Eq.~(\ref{eq2}) at $m_{c\bar{c}}^2=4M_c^2$ and $z=1$.
In the NRQCD model, the produced $c\bar{c}$ pairs with large
$m_{c\bar{c}}^2$ have very small probabilities to become J/$\psi$
mesons.  On the other hand, the CE model assumes that the
$D_{[c\bar{c}]}$ are independent of the pairs' invariant masses
(provided they are below the open charm threshold), and are not
sensitive to the pairs' quantum numbers.  We can represent the CE model
by taking $D_{[c\bar{c}]}=f_{{\rm J/}\psi} \delta(1-z) 
\theta(4M_D^2-m_{c\bar{c}}^2)$ in Eq.~(\ref{eq2}) with a fitting
constant $f_{{\rm J/}\psi}$ independent of the $c\bar{c}$ pairs'
quantum numbers.  

Since the NRQCD and the CE models are both consistent with the
existing data on {\it inclusive} J/$\psi$ production, the new QCD
factorized formula should also be consistent with the data
\cite{QS-jpsi}.  However, both models have difficulties explaining the
recent CDF data from Fermilab on J/$\psi$ polarization at large
transverse momentum \cite{CDF-jpsi}.  

\begin{figure}
\begin{center}
\begin{minipage}[t]{5.9in}
\begin{center}
\epsfig{figure=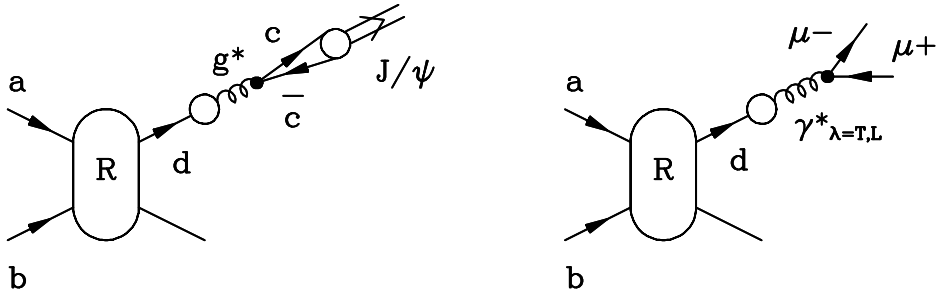,width=5.2in}
\vskip -0.35in
\caption{Sketch for J/$\psi$ and Drell-Yan production at large
transverse momentum}
\end{center}
\end{minipage}
\end{center}
\vspace{-0.25in}
\end{figure} 

At large transverse momentum, J/$\psi$ are produced mainly from the
fragmentation contribution, as shown in Fig.~1, where a
virtual gluon is produced and decays into a $c\bar{c}$
pair which transmutes into a physical J/$\psi$.  The polarization of
the produced J/$\psi$ depends on the polarization of the 
$c\bar{c}$ pair and the details of the transition distributions 
$D_{[c\bar{c}]}$.  In order to use the polarization measurements
to discern between the production models, we need independent tests of
the reliability of the QCD formalism for calculating the $c\bar{c}$
polarization at large transverse momentum.  As illustrated in
Fig.~1, the low mass Drell-Yan lepton-pair's angular 
distributions at large transverse momentum have a lot in common with
the $c\bar{c}$ polarization \cite{QRZ-alpha}.  If the QCD calculations
are consistent with the virtual photon polarization in Drell-Yan
massive lepton-pair production, we can 
conclude that the QCD formalism for calculating the polarization of
the $c\bar{c}$ at large transverse momentum is reliable, and
therefore, measurements of J/$\psi$ polarization provide 
decisive tests for models of J/$\psi$ production. 

\section{J/$\psi$ Suppression}

In addition to polarization measurements, the nuclear medium
dependence of J/$\psi$ production is sensitive to the J/$\psi$
formation process \cite{QVZ-jpsi}.  If the J/$\psi$ mesons are formed
immediately after the production of the $c\bar{c}$ pair, the medium
dependence is dominated by the hadronic scattering between a
color-singlet J/$\psi$ and the nuclear matter.  On the other hand, if
the J/$\psi$ mesons are formed much later, the observed medium
dependence is a consequence of the dynamical interactions between the
pre-J/$\psi$ $c\bar{c}$ states and the nuclear medium.  Therefore,
the nuclear dependence can help select the correct model for
J/$\psi$ production.

In proton-nucleus or nucleus-nucleus collisions, the produced
$c\bar{c}$ pairs are likely to interact with the nuclear medium before
they exit. If we assume each interaction between the $c\bar{c}$ pair
and the nuclear medium is about the {\it same} and can be
treated {\it independently}, we naturally derive the Glauber
formula: $\sigma_{AB}={\rm exp}[-\rho_N \sigma_{\rm abs} L(A,B)]$ with
effective absorption cross section $\sigma_{\rm abs}$ to change a
$c\bar{c}$ pair to a pair of open charms, and effective length
$L(A,B)$ of the medium in A+B collisions.  The characteristic feature
of the Glauber formula is a straight line on a semi-log plot of
$\sigma_{AB}$ vs. $L(A,B)$, which is not consistent with the strong
J/$\psi$ suppression observed in Pb-Pb collisions \cite{NA50-jpsi}.

Our model of J/$\psi$ suppression is based on the following points:
(1) it is the $c\bar{c}$ pair, not the J/$\psi$ meson, that interacts
through most of the nuclear medium; (2) multiple interactions of the
$c$ and/or $\bar{c}$ with soft gluons in the medium increase the
relative momentum between the $c$ and $\bar{c}$; and (3) as the
relative momentum increases, the phase space to form a J/$\psi$ meson 
decreases \cite{QVZ-jpsi}.  Since multiple soft-scatterings lead to a
larger relative momentum between the $c$ and $\bar{c}$, the effective
$\sigma_{\rm abs}$ of the pair increases as the $c\bar{c}$ passes
through the medium leading to a stronger suppression than given by
the simple Glauber formula which applies to single particle
propagation \cite{QVZ-jpsi}.

Integrating over J/$\psi$'s momentum in Eq.~(\ref{eq1}), we obtain
a factorized expression for the J/$\psi$ total cross section
\begin{equation}
 \sigma_{AB\rightarrow {\rm J/}\psi}
\approx \sum_{a,b,[c\bar{c}]}
 \int dx \phi_{a/A}(x)\int dx' \phi_{b/B}(x') \int dm_{c\bar{c}}^2
 \frac{d\hat{\sigma}_{ab\rightarrow [c\bar{c}]}}
      {dm_{c\bar{c}}^2}
 F_{[c\bar{c}]\rightarrow{\rm J/}\psi}(m_{c\bar{c}}^2)
\label{eq3}
\end{equation}
where $F_{[c\bar{c}]\rightarrow{\rm J/}\psi}$ represents an inclusive
transition probability for a $c\bar{c}$ pair of invariant mass
$m_{c\bar{c}}^2$ to become a J/$\psi$ meson.  
Without the large transverse momentum, the corrections to
Eq.~(\ref{eq3}) from the spectator interactions, which are suppressed
by a factor $\sim \rho_N L(A,B)/4M_c^2$, become more important because
of the large medium length $L(A,B)$.  Just like the random walk of two
particles, the soft spectator interactions effectively increase the
invariant masses of the $c\bar{c}$ pairs, and therefore, reduce the
phase space to form J/$\psi$ mesons.  We calculate the J/$\psi$
suppression in the nuclear medium by including the 
multiple spectator interactions to the pre-J/$\psi$ $c\bar{c}$ pairs,
and plot our results in Figs.~2 and 3.  The data are from
Ref.~\cite{NA50-jpsi}.  The dashed lines are predictions of the
Glauber formula, and the solid lines are the results of our
calculations.  With only one parameter $\varepsilon^2$, which is
defined as average gain of invariant mass per unit medium length and
is fixed by the total suppression in Fig.~2, our calculations of the
J/$\psi$ suppression are consistent with all existing data, except a
couple of points at the highest $E_T$ bins of the New NA50 data
\cite{QVZ-jpsi}. 

\begin{figure}
\begin{center}
\begin{minipage}[t]{3.0in}
\begin{center}
\epsfig{figure=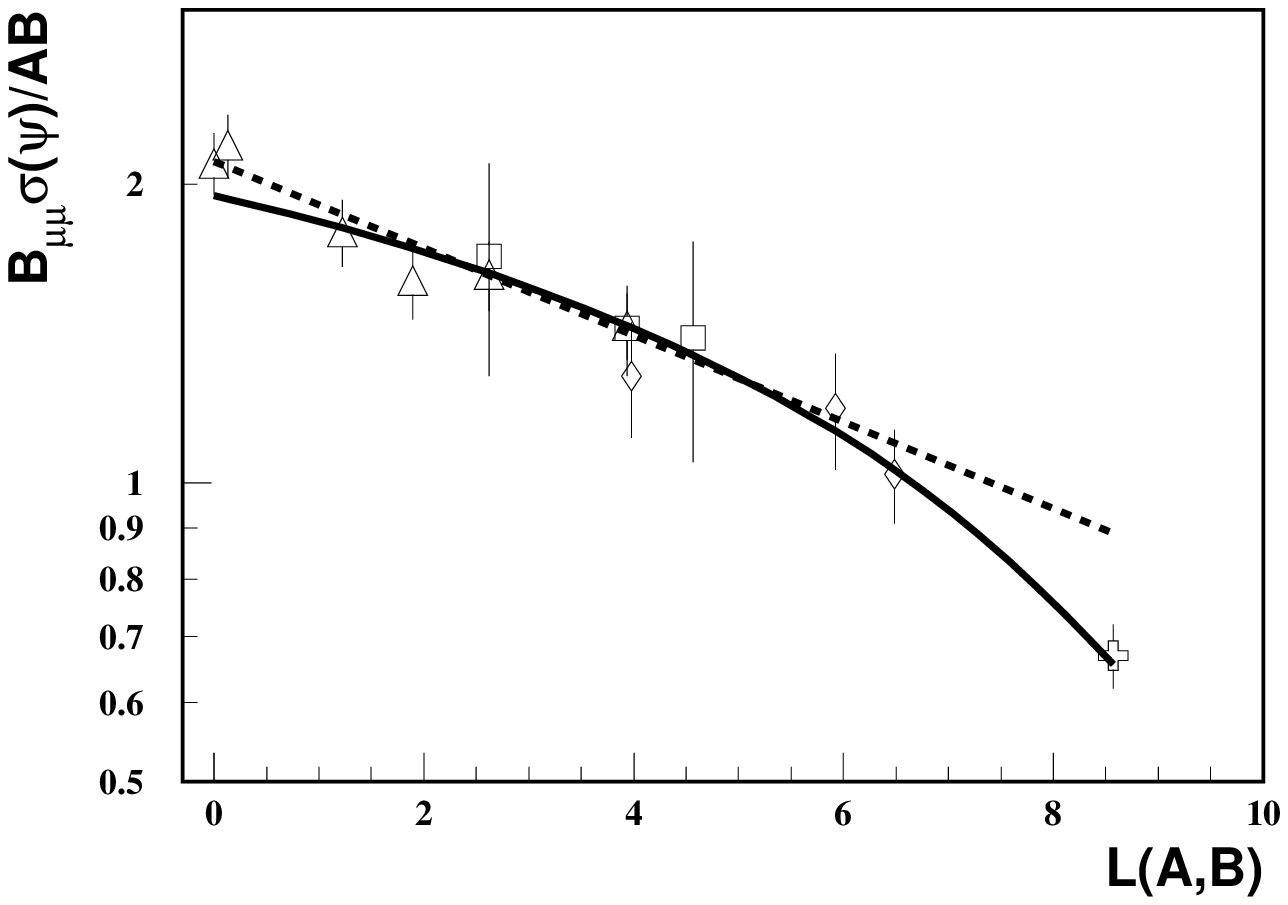,width=3.0in}
\vskip -0.4in
\caption{J/$\psi$ cross section to $\mu^+\mu^-$}
\end{center}
\end{minipage}
\hfil
\begin{minipage}[t]{3.0in}
\begin{center}
\epsfig{figure=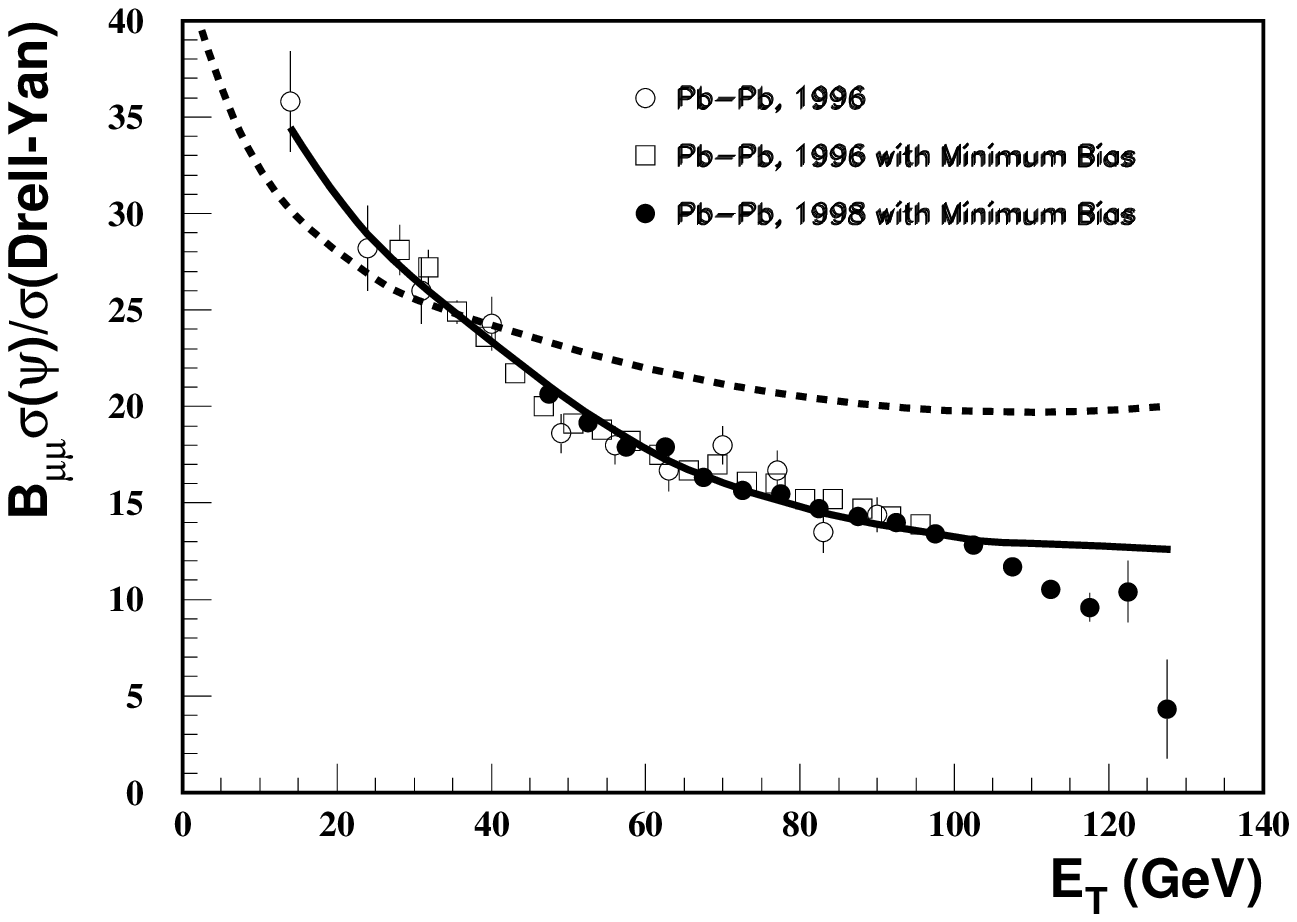,width=3.0in}
\vskip -0.4in
\caption{Ratio of J/$\psi$ over Drell-Yan}
\end{center}
\end{minipage}
\end{center}
\vskip -0.25in
\end{figure} 

We conclude that the NRQCD model and the CE model are closely
connected to the new QCD factorization formula for J/$\psi$
production.  The measurements of J/$\psi$ polarization could provide a
decisive test of the production mechanism.  Our model of J/$\psi$
suppression is consistent with nearly all existing data on J/$\psi$
total cross section. 

This work was supported in part by the U.S. Department of
Energy under Grant No. DE-FG02-87ER40731.

\end{document}